\documentstyle[prl,twocolumn,aps,epsfig]{revtex}

\begin{document}
\draft
\title{Structure of Vortices in Two-component Bose-Einstein Condensates}
\author{D. M. Jezek$^{1,2}$, P. Capuzzi$^{1}$, and H. M. Cataldo $^{1,2}$.}
\address{$^1$Departamento de F\'{\i}sica, Facultad de Ciencias Exactas
y Naturales, \\
Universidad de Buenos Aires, RA-1428 Buenos Aires, Argentina}
\address{$^2$Consejo Nacional de Investigaciones Cient\'{\i}ficas y T\'ecnicas,
Argentina}

\maketitle

\begin{abstract}

We develop a three-dimensional analysis of the phase separation of
two-species Bose-Einstein  condensates in the presence of
vorticity
 within the Thomas-Fermi approximation.
We find  different segregation features according to whether the
more   repulsive component is in a vortex  or in a vortex-free
state. An  application of this study is aimed at describing
systems formed by two almost immiscible species of rubidium-87
that are commonly used in Bose-Einstein condensation experiments.
 In particular, in this work we calculate the density profiles
of condensates for the  same  conditions as
 the states prepared  in  the experiments
performed at JILA [ Matthews, {\it et al}., Phys. Rev. Lett. {\bf
83},
 2498 (1999)].

\end{abstract}

\pacs{03.75. Fi, 67.57.Fg, 67.90.+z }

Vortices in  Bose-Einstein condensates (BECs) have thoroughly been
studied from a theoretical point of view since the first
experiments on condensates were carried out, due to their direct
connection to superfluidity. For a review of these issues see for
example \cite{da99}, and
 for the most recent works references in  \cite{todos}.
But until the experiment of Matthews {\it et al.} \cite{ma99} was
performed, no evidence about their existence had ever been
 reported.
This experiment was based on a  method proposed by Williams and Holland
\cite{wi99}  to create vortices in  two-component BECs.
They trapped atoms of $^{87}$Rb in two different hyperfine
states $ |F=1, m_f=-1\rangle $ and $ |F=2, m_f=1\rangle $,
henceforth denoted as
 $ |1\rangle $ and
 $ |2\rangle $, respectively, and created  vortices either 
in the $ |1\rangle $ or
$ |2\rangle $ component, while the other species remained
without rotation. Because of the relation between intra- and
inter-particle scattering lengths, these states have  the
property of being almost immiscible. We shall see that this fact
has important consequences on the structure of the vortices.

In connection with  this  experiment interesting theoretical work
has recently been developed  \cite{ga00,pe99,chu00} in order to
describe different properties of many types of two-component
condensate systems. Garc\'{\i}a-Ripoll and  P\'erez Garc\'{\i}a,
by means of a  robust formalism, analyzed the stability and
dynamics of a variety of two-species condensates, taking into
account the effect of the interchange of
 species in the system, and also the  variation of the number of particles.
 In particular, in Ref. \cite{ga00} they set
the relative number of particles equal in both species and varied
the total number of particles, and in Ref. \cite{pe99} they
pursued the research by also varying the relative composition
using a 2D model. 
However, in these works \cite{ga00,pe99} little is said about the
structure of the stationary states and no comments are made about
segregation of species. In a recent work, Chui {\it et al}.
\cite{chu00} computed the energy
 of different configurations within the
Thomas-Fermi Approximation (TFA). They varied the
number of particles and interchanged the species in order
to derive   conclusions about the relative stability.
But, although they  used a 3D model and considered
 phase separation in the mixture,
a  hypothesis  was made
regarding the form of the interface between species that we find is not
consistent  with our results and apparently also with new available 
experimental data \cite{and00}. We shall comment on this later.

In 1996, in a pioneering work,  Ho and  Shenoy  \cite{ho96}
constructed an algorithm to determine the density profiles of 
binary  mixtures of alkali atoms within the TFA.
We develop here a different procedure which allows one to obtain
conclusions about the general features of segregation through simple
 calculations.

The aim of the present article is to analyze and perform a
classification of  the structure of condensates near phase
separation, when two almost immiscible species are involved. Our
starting point for calculating the stationary states are the
Gross-Pitaevskii equations in the TFA \cite{da99}:
\begin{eqnarray}
\mu_1 \Psi_1 & = & (V_1 + N_1 G_{1,1} |\Psi_1|^2 +
 N_2 G_{1,2}  |\Psi_2|^2) \, \Psi_1, \nonumber \\
\mu_2 \Psi_2 & = & ( V_2 + N_1 G_{1,2} |\Psi_1|^2 +
 N_2 G_{2,2}  |\Psi_2|^2) \, \Psi_2,
\label{tfa}
\end{eqnarray}
where $ N_i$ denotes the number of atoms of species $i$, and
$G_{k,l}= u_{k,l} U $, with  $ u_{k,l} $   the relative
interaction strengths between species $k$ and $l$. We set the
most repulsive component $|1\rangle$  in the  $ \Psi_1 $ state,
and fix $ u_{1,1}= 1$ $ (> u_{2,2}) $, so $ U= \frac{4 \pi
\hbar^2 a }{M} $,
 $ a $ being the scattering length of the $|1\rangle$-species and
$M$  the atom mass.

For simplicity, we consider the system in a  spherically symmetric
trap and make a change of  variables according to
$\sqrt{\frac{M}{2}} w_0 \vec{r} \rightarrow   \vec{r} $, with $
w_0$ the trap angular frequency. Thus  the effective potentials
$V_i$ written in cylindrical variables read
\begin{equation}
V_{i} =  r^2 +  z^2 + \frac{\kappa_i}{r^2}.
\label{pot}
\end{equation}
The first two terms correspond to the trapping potential, while
the third  includes the centrifugal term  \cite{ho96}, $ \kappa_i
= \frac{m_i^2 \hbar^2 w_0^2}{4} $, with $m_i$ the number of quanta
of circulation associated with the wave function for species $i$.

Let us first analyze  what happens when
 $D \equiv u_{1,1} u_{2,2}- u_{1,2}^2 =0 $. Physically the system starts
undergoing phase separation. Mathematically, for both $ |\Psi_i|^2
> 0$,
 Eqs. (\ref{tfa}) become  linearly dependent. So to have a compatible
set of equations the following condition must hold:
\begin{equation}
  \frac{ \mu_1}{G_{1,1}}-\frac{ V_1}{G_{1,1}}=
 \frac{ \mu_2}{G_{1,2}}-\frac{ V_2}{G_{1,2}}.
\label{sur}
\end{equation}

If we define $ \beta_2 = u_{1,2}/u_{1,1} $, the above equation can
be rewritten as

\begin{equation}
 \frac{ \mu_2 -\beta_2 \mu_1}{1-\beta_2} - r^2 - z^2
+\frac{\beta_2 \kappa_1 - \kappa_2}
{(1-\beta_2) r^2} = 0.
\label{sup}
\end{equation}
This equation defines different families of surfaces that we shall
call $ \cal{S}$$_0^i$, where $i$ labels the family. As was
mentioned above, this is the only region of space where both $
\Psi_i $ may be non-zero, so these surfaces  necessarily determine
the interfaces between species. Such families are characterized by
the values of

\begin{equation}
A= \frac{ \mu_2 -\beta_2 \mu_1}{1-\beta_2} , \,\,\,\,\,
B= \frac{ \beta_2  \kappa_1 -\kappa_2}{(1-\beta_2) },
\label{a,b}
\end{equation}
(note that $ D=0 $ implies
 $ \beta_2 <1 $).

Depending on the values of the above parameters, three
topologically different kinds of surfaces exist, namely,
spheres, cylinder-like surfaces, and  toroids. In Fig. 1 we show
a partition of the parameter space $(A,B)$ indicating the
domains corresponding to each family. In order to classify these
surfaces let us study the $ z(r) $ curves defined by Eq.
(\ref{sup}). If $ z(r) $ intersects the $z$ axis, the associated
surfaces $
\cal{S}$$_0^a$ are spheres, and this occurs for  $ B=0 $ and $ A>0
$.
 If $z(r)$ has only one root  the corresponding surfaces
 $ \cal{S}$$_0^b$ are cylinder-like, and this takes place when  $ B>0 $.
Finally, if $z(r)$ has two roots  the related surfaces $
\cal{S}$$_0^c$ are  toroids, and it is easy to verify that  the
conditions to be fulfilled
 are $ A^2+4B > 0 $, $A>0$, and $ B<0 $.
 The remaining region of the parameter space has no associated surfaces.

To perform a displacement of a point within the parameter space two
mechanisms are possible.
 On the one hand, the  $ A$  value may be
changed by varying the number of particles  in each species, which
will cause a change in the chemical potentials. A variation of
this type does not change the family, but the shape of the surface
is modified. For example, for $ B>0 $ one can get a straighter
``cylinder" by  moving  from positive to negative $A$ values, and,
in terms of the number of particles, this corresponds to a
decrease in the $N_2/N_1$ ratio. On the other hand, $B$ can only
take discrete values and it is modified by changing  the
vorticity.

It is  worthwhile  mentioning  that the only possible realization
for  $B=0$, (and hence for $ \cal{S}$$_0^a$), within reasonable
values of $ m_i$,  is a vorticity free condensate $m_1=m_2 =0$.
In such a case, regarding the location of each species,
it is easy to conclude that the energy minimum is obtained
when the less repulsive component  lies  inside the sphere of radius
 $ \sqrt{A} $, while the other species forms a spherical shell around it.
In the presence of vorticity ($B \ne 0 $), the location of each
species can be predicted from energetic considerations. The state
with greater vorticity  should be located in the region of space
that excludes $ r=0 $, because of the centrifugal term.

From this simple analysis we may state that close to phase separation
the presence of vorticity dramatically changes the segregation structure.

Now, let us consider the general case  $ D \ne 0 $. The solution
of Eqs. (\ref{tfa} ) can  be easily  obtained and has the
following different expressions depending on whether there exists
any overlap between the wave functions of both species. a) In the
region where   only one wave function is non-vanishing ( $
|\Psi_{i}|^2 \ne 0 $   and $ |\Psi_{k}|^2 = 0 $, for $ i \ne k$)
Eqs. (\ref{tfa}) are decoupled and the solution is
\begin{equation}
 |\Psi_{i}|^2=  \left[ \mu_i -r^2-z^2-\kappa_i/r^2 \right]/(G_{i,i} N_i).
\label{fui}
\end{equation}
b) In the region where both wave functions are non-vanishing,
 $  |\Psi_{i}|^2 > 0 $, one obtains

\begin{eqnarray}
|\Psi_{1}|^2 & = & \left[ \frac{\mu_1- \beta_1 \mu_2}{(1 - \beta_1)} - r^2
-z^2+ \frac{\kappa_2 \beta_1
-\kappa_1}{r^2 (1 - \beta_1)} \right]  A_1 \,  u_{2,2}, \nonumber \\
 |\Psi_{2}|^2 & = & \left[ \frac{\mu_2- \beta_2 \mu_1}{ (1- \beta_2 )}- r^2
-z^2+  \frac{\kappa_1 \beta_2
-\kappa_2}{r^2 (1 - \beta_2)}\right] A_2 \, u_{1,1},
\label{fu}
\end{eqnarray}
where  $ \beta_1 = \frac{u_{1,2}}{ u_{2,2}}$ and $ A_i = \frac{ (1
- \beta_i) }{U N_i D} $.

We call  $ \cal{S}$$_1$ ( $ \cal{S}$$_2$) the surface defined by
equating  the expression inside the square bracket in
$|\Psi_{1}|^2$ ($ |\Psi_{2}|^2$) to zero. It is interesting to
notice that the expressions mentioned above are of the same type
as the one that defines $ \cal{S}$$_0$ (cf. Eq. (\ref{sup})),
and so a similar  classification can be applied to these
surfaces. The surfaces $ \cal{S}$$_1$ and $ \cal{S}$$_2$ define
the boundary of the coexisting region  of the $ |1\rangle $ and
$ |2\rangle $ species. Depending on the value of the
interparticle interaction strength, the regions where the
overlapping of the waves function occurs exhibit different
features. For $ D > 0 $ and  $ u_{1,2} $  small enough, the
coexisting region can be very large, as may be observed in the
figures shown in Ref.\cite{ho96}. When increasing $ u_{1,2} $
this region becomes narrower. At $D=0$, it is easy to verify
that $ \cal{S}$$_1$ $ = $ $ \cal{S}$$_2$ $ = $  $
\cal{S}$$_0$, and hence the  coexisting region is reduced to a
single surface. When $ u_{1,2}$ increases making $D<0$,  this
surface splits into two distinct surfaces, forming  thus a new
narrow 3D overlapping region, due to mutual repulsion of the
species.

To have  the problem completely solved one has to calculate the
chemical potentials. In order to  obtain  $ \mu_1 $ and  $ \mu_2$
the normalization condition for the wave functions given by Eqs.
(\ref{fui}) and (\ref{fu}) must be used. This leads to a system of
two coupled  integral equations that must be solved  numerically.
We have accomplished this in order to describe the states obtained
in the JILA experiment \cite{ma99}, choosing parameters equal to
those of the experiment. Namely, the trapping potential has an
angular frequency
 $  w_{0} = 2 \pi \, \nu_{\text{trap}}$,
 being $ \nu_{\text{trap}}= 7.8 $ Hz. 
The interaction strengths relative to the
$ |1\rangle $  component of the   $^{87}$Rb species are
$u_{1,1}=1,u_{1,2}= .97 $ and $u_{2,2}= .94 $. And the number of
particles of each species is $N_1= N_2 = 4$$ \times$$ 10^5 $.
Note that for these values of the interaction strengths, $ D=
-9$$
\times$$10^{-4} $ is close to the phase separation value, so the
region between interfaces is very narrow, and hence $ \cal{S}$$_1$
$\sim$  $ \cal{S}$$_2$  $\sim$  $ \cal{S}$$_0$. Thus hereafter, in
order to identify the type  of segregation of species encountered,
we will make use of the classification we performed for  $
\cal{S}$$_0$.

For completeness, we  also include calculations for the two
species without vorticity, although the experimental realization
consisted only of one vortex and one non-vortex state in each
component.

In Fig. 2   we show the density contours of each wave function in
the (r,z) plane for the following three configurations.
(i) A vortex-free system $ (m_1,m_2) =(0,0) $ yields chemical
potentials in units of  $ \hbar  w_{0} $  as $ \mu_1=24.50 $ and
 $ \mu_2=24.06 $. It may be seen that in this case the
segregation evolves as  $\sim$  $ \cal{S}$$_0^a$. The more
repulsive component is  on a spherical shell around the other, as
expected. (ii) For  $ (m_1,m_2) = (1,0) $  we obtained $
\mu_1=24.54 $ and $ \mu_2=24.04 $. The vortex segregates in a ring
surrounding the non vortex state. It is easy to verify that the
species become separated by  an interface of the type $
\cal{S}$$_0^b$. The core size is about $ 9 \mu$m and the whole
vortex has a radius of  $ 26 \mu$m.  This reproduces very well the
experimental data. (iii) When the vortex is in  the less repulsive
component $ (m_1,m_2) =(0,1) $, we obtained $ \mu_1=24.50 $ and $
\mu_2=24.15 $. In this case the vortex  is segregated inside the
non-vortex state. The corresponding surface that describes such
segregation is  the toroid $ \cal{S}$$_0^c$. The inner and outer
radii of the torus  are  $ 5 \mu$m and  $ 18 \mu$m,
respectively. However, this does not correspond to what was
obtained at the first stage of the experiment,  which looks
quite similar to the pattern described in (ii). Mainly, in the
experiment, the radius of the vortex was approximately  $ 26 \mu$m,
 while according to  our calculations it should be  
 $18 \mu$m. Taking into account that after vortex creation, the
vortex immediately shrank, a possible and likely explanation for
the discrepancy is that the system was not created near the
equilibrium configuration but moved towards it. The case with a
vorticity of $ (m_1,m_2) = (1,0) $ is created much closer to
equilibrium \cite{a00}.  It is important to notice that in every case, 
even when the vortex configuration is out of equilibrium, the whole
condensate is almost spherical with the same radius value of $
26
\mu$m.

In Fig. 3 we display the density profiles as a function of the
radial coordinate at $z=0$, for the same three configurations
described in Fig. 2. It may be seen in all of the  cases that the
shape of the total density is like a parabola, a characteristic
that also holds for any other direction considered. Regarding the
densities of each one of the components separately, they abruptly
go to zero in the narrow region between species.
It is worthwhile to mention that
this behavior would not be so sharp for  real systems and some 
 interpenetrating region would exist \cite{ha98}. 

With respect to the hypothesis made  by  Chui { \it et al}.
\cite{chu00}, they
 assumed  for condensates including vorticity  
in  a symmetric trap that one component
is confined inside a ball and the other one in a spherical shell
around it; and that  the species with vorticity have a small healing
length sized vortex core.
 In the context of
our article this means that they worked as if segregation of
species, including vortices, always evolves as  $\sim$  $
\cal{S}$$_0^a$, which is not consistent  with our
results.
 In a very recent experimental work \cite{and00} Anderson { \it et al}.
have included  a figure (Fig. 3) in which a 
two-component vortex is 
viewed along its axis direction and  along an orthogonal one.
Views of   a vorticity free condensate are also included.
The in-trap pictures of Ref. \cite{and00} cannot rule out a
healing length core  because such a 
feature would be too small ( $ \sim 0.7 \mu$m ) to resolve. 
So in this experiment Chui's condensates
should look like a vortex free condensate. As it may be seen 
in the above mentioned figure  the images of a system 
with and without vorticity are
quite different.
In particular, we want to note that in the vortex case there is a  visible
evidence of a cylinder type segregation.
However to be conclusive about these statements more experimental 
information is required.

In Refs. \cite{ga00,pe99} the authors employed, for  describing
the stationary states, wave functions expanded on a finite subset
of the harmonic oscillator basis. This means that they did not
work with  exact solutions; unfortunately in these articles there
is no precise information about the structure of the states.
 However, looking at both
the domains of the wave function (Fig. 2) and the form of the
density itself (Fig. 3), it seems that they should have taken into
account a  large number of elements of this basis
 to describe similar configurations.

It is well known that  the TFA turns out to be very  accurate
when systems with a large number of particles
are considered, so the main features of our analysis could be regarded
as a guide to check the accuracy of different approximations.

Finally, we want  to mention that very interesting experimental
\cite{an00} and theoretical  \cite{mc00} work dealing with
 off-centered vortices in two-species condensates, is forthcoming.

In conclusion, we think that our  analysis of states within the
TFA could be helpful  both from the experimental and theoretical
viewpoints. In the first case, to prepare a configuration as close
as possible to a stationary state, and in the second, to make a
suitable choice for the set of wave functions  in which such kinds
of states are to be expanded.

We gratefully acknowledge useful conversations with B. P. Anderson.

\begin{figure}
\centering
\epsfig{file=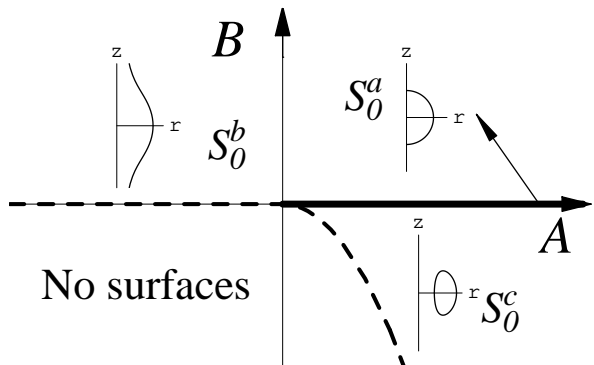,clip=,width=.9\columnwidth}
\caption{ Domains of the different families of surfaces $
\cal{S}$$_0^i$ in parameter space $ (A,B) $. In each region we
indicate the  related $z(r)$ curve. The thick line $B=0$ and
$A>0$ corresponds to  spheres. The $B>0$ region corresponds to
cylinder-like surfaces. And for the conditions $A>0$, $B<0$, and
for $ B > - \frac{A^2}{4}$ the surfaces are toroids. The dashed
line determines the separation between these families of
surfaces and empty sets.} \label{fig1}
\end{figure}
\begin{figure}
\centering
\epsfig{file=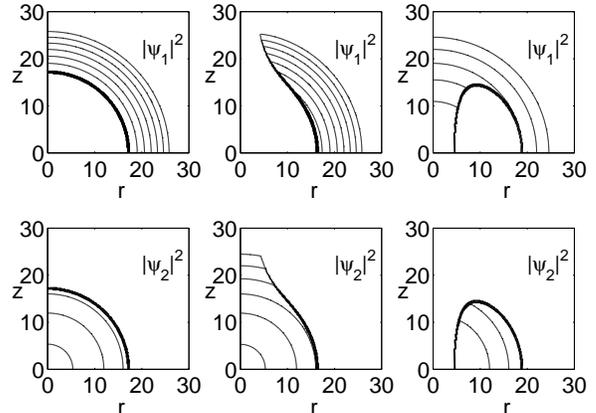,clip=,width=.9\columnwidth}
\caption{ Density contours in the (r,z) plane for  component $
|1\rangle$ ($|2\rangle$) in the first (second) row of
subfigures. The columns correspond to the following situations:
(left) no vortex $ (m_1,m_2)=(0,0) $, (middle) a vortex in $
|1\rangle $ species $ (m_1,m_2)=(1,0) $, and (right) a vortex in
the less repulsive component $ (m_1,m_2)=(0,1) $. The density
contour spacing is $ 10^{-5}$ $  \mu$m$^{-3}$. }
\label{fig2}
\end{figure}
\begin{figure}
\centering
\epsfig{file=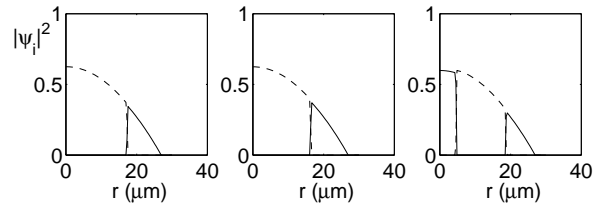,clip=,width=.9\columnwidth}

\caption{ Density profiles $ |\Psi_1|^2 $ (solid line) and
$ |\Psi_2|^2 $ (dashed line) as function of $ r $ at $ z=0 $
for the same three distributions of vorticity described in the
previous figure. The density is given in units of
 $ 10^{-4}$ $  \mu$m$^{-3}$. }
\label{fig3}
\end{figure}
\end{document}